\documentclass{icrc2009}

\usepackage{graphicx}   
\usepackage[caption=false]{caption}    
\usepackage[font=footnotesize]{subfig} 
\usepackage{fixltx2e}
\usepackage{url}

\newcommand{\shorttitle}[1]%
{\markboth{Proceedings of the 31\MakeLowercase{$^{st}$} ICRC, {\L}\'{o}d\'{z} 2009}{#1} }
\newcommand{\etal}{\MakeLowercase{\textit{et al. }}} 

\begin{document}
\title{Microquasar observations with the MAGIC telescope}

\author{\IEEEauthorblockN{T.~Y.~Saito\IEEEauthorrefmark{1},
			  R.~Zanin\IEEEauthorrefmark{2},
                          P.~Bordas\IEEEauthorrefmark{3}, 
                          V. Bosch-Ramon\IEEEauthorrefmark{3},
                          T. Jogler\IEEEauthorrefmark{1},
                          J. M. Paredes\IEEEauthorrefmark{3},
                          M. Rib\'{o}\IEEEauthorrefmark{3},\\
                          M. Rissi\IEEEauthorrefmark{4},
                          J. Rico\IEEEauthorrefmark{5},\IEEEauthorrefmark{2} and
                          D. F. Torres\IEEEauthorrefmark{6}
                          for the MAGIC Collaboration}\\
\IEEEauthorblockA{\IEEEauthorrefmark{1}Max-Planck-Institut f\"ur Physik, D-80805 M\"unchen, Germany}
\IEEEauthorblockA{\IEEEauthorrefmark{2}IFAE, Edifici Cn., Campus UAB, E-08193 Bellaterra, Spain}
\IEEEauthorblockA{\IEEEauthorrefmark{3}Universitat de Barcelona (ICC/IEEC), E-08028 Barcelona, Spain}
\IEEEauthorblockA{\IEEEauthorrefmark{4}ETH Zurich, Switzerland}
\IEEEauthorblockA{\IEEEauthorrefmark{5}ICREA, E-08010 Barcelona, Spain}
\IEEEauthorblockA{\IEEEauthorrefmark{6}Institut de Ciencies de l'Espai (IEEC-CSIC), Bellaterra, Spain} 
}

\shorttitle{T.~Y. Saito \etal Microquasar observations with MAGIC}
\maketitle

\begin{abstract}

Microquasars, X-ray binaries displaying relativistic jets driven by accretion onto a compact object, are some of the most efficient accelerators in the Galaxy. Theoretical models predict Very High Energy (VHE) emission at the base of the jet where particles are accelerated to multi-TeV energies. This emission could be detected by present IACTs. 
The MAGIC telescope observed the microquasars GRS~1915+105, Cyg X-3, Cyg X-1 and SS433 for $\sim 150$ hours in total from 2005 to 2008. We triggered our observations by using multi wavelength information through radio flaring alerts with the RATAN telescope as well as by ensuring the {\it low/hard} state of the source through RXTE/ASM and {\it Swift}/BAT monitoring data. 
We report on the upper limits on steady and variable emission from these sources over this long period.
\end{abstract}


 \section{Introduction}

 Microquasars (MQ) are X-ray binary systems containing neutron stars or black holes that display relativistic jets \cite{Mirabel1999}. Currently $\sim$~16 of these sources are known in our Galaxy \cite{Ribo2005}. Since their discovery, MQs have attracted the attention of the high energy astrophysics community because they are scaled-down versions of quasars, and allow the study of jet emission and formation on much shorter timescales. 
 An evidence of TeV emission has been obtained in the stellar-mass black hole
MQ Cyg X-1 \cite{Albert2007}. A MQ scenario \cite{Paredes2000} \cite{Massi2004}  has also been suggested for  the
recently detected TeV sources LS 5039 \cite{AharonianHESS2005a} and  LS I +61 303 \cite{Albert2006}
though they could otherwise contain a young non-accreting pulsar and be powered by its rotation \cite{Dubus2006} \cite{Dhawan2006}, as is the case for PSR B1259 -63 \cite{AharonianHESS2005b}.
The detection of new MQs in the TeV domain is therefore an important target for the present VHE gamma-ray observatories. 

We present the MAGIC observations of four well-established galactic MQs: GRS~1915+105, Cyg~X-1, Cyg~X-3 and SS~433. These sources present unique properties concerning the non-thermal activity, the kinetic power present in their jets and the surrounding radiation/matter fields where gamma-rays could be generated both in leptonic and hadronic scenarios. The observations span four cycles of operations of the telescope from 2005 to 2008, resulting in a variety of physical conditions (regarding both orbital phases and spectral states) in which VHE emission could be produced. 

In Sect.~2, we refer the MAGIC telescope performance, the data reduction method and analysis technique applied. In Sect.~3 we present the data sample and the trigger strategy actually used in our observations. We report the resulting upper limits to the TeV emission for each source, both for the whole sample and in a phase-folded basis. in Sect.~4. Day-by-day analysis for transient emissions is also shown. Sect.~5 is devoted to the discussion of the obtained results.

\section{The MAGIC Telescope and \\Analysis Technique} \label{Technique}

The MAGIC  telescope \cite{MAGIC1} is a new generation Imaging Atmospheric Cherenkov Telescope at La Palma,
Canary Islands, Spain (28.3$^{\circ}$N, 17.8$^{\circ}$W, 2240 m a.s.l.), successfully operating since
 the beginning of 2004.
The parabolically-shaped reflector,
with its total mirror area of 236 m$^2$, allows MAGIC to sample a
part of the Cherenkov light pool and focus it onto a multi-pixel
camera, composed of 576 photomultipliers (PMTs). 

The output pulses of PMTs are converted into optical signals,
transmitted via optical fibers and digitized by a Flash Analogic to Digital Converter (FADC) System.
In the first years of the MAGIC operations, 300 MSample/s FADCs (Siegen-FADCs) were used and it was upgraded to 2 GSample/s ultra-fast FADCs (Mux-FADCs) in February 2007, reducing contamination of night sky background and improving the reconstruction of the timing characteristics of the recorded images.

The data analysis was performed using the standard MAGIC analysis
software.
After the calibration \cite{MARS} and the image cleaning \cite{timeimagecleaning}, the shower images were parametrized by the so-called Hillas image parameters \cite{Hillas}. 
In the case of data taken with the Mux-FADCs, two additional parameters, namely the time 
gradient along the main shower axis and the time spread of the shower pixels, were also computed \cite{timeimagecleaning}.
Hadronic background suppression was achieved
using the Random Forest (RF) method \cite{RF}
where for each event the so-called ``hadronness''
was computed based on the Hillas and the time parameters. 
``Hadronness'' is a measure of the probability that the event is not gamma
 like. The RF method was also used for the energy estimation. 
 The calculation of the number of excess events was performed in two different ways
in the data before and after the installation of the Mux-FADCs: in Siegen-FADC data, DISP method \cite{disp2} was used, while Alpha method \cite{timeimagecleaning} was used in Mux-FADC data
so that the timing parameters could have been used with less systematic uncertainties. 
 Data taken with significantly lower or higher trigger rate and with zenith angle larger than 50$^\circ$ were rejected from the analysis. 
 Images with less than 200 ph.e. were discarded in order to assure the better background rejection and to avoid a systematic error in estimation of the background.
 The sensitivity of the telescope is estimated to be $\sim~2\%$ and 1.6\% C.U. for 5$\sigma$ detection in 50 hours of low zenith observations, with Siegen-FADCs and Mux-FADC, respectively \cite{timeimagecleaning} \cite{Crab}.


\section{Trigger Strategy and Data Sample} \label{Data}

In 2005 and 2006, the MAGIC telescope observed GRS 1915+105 and Cyg X-3 for 14 hours each. These observations were triggered by the alerts on the source flaring state at radio frequencies sent by the RATAN-600 telescope.
On the other hand Cyg X-1 was monitored between July and November 2006 for 46 hours and an evidence of signal at the level of 4.0 $\sigma$ (pre-trial) was observed on September 24th, 2006 \cite{Albert2007}, when the source was in a {\it low/hard} state and in coincidence with a hard X-ray flare.
 Following this promising result, in 2007, we moved to a monitoring approach when the sources stayed in the {\it low/hard} state, characterized by the presence of steady relativistic jets where VHE emission could be produced. The spectral state was defined by using public soft and hard X-ray data from All Sky Monitor (ASM) \cite{ASM}, hosted by the {\it Rossi} X-ray Timing Explorer (RXTE) satellite, and from the Burst Alert Telescope (BAT) \cite{BAT} on board of {\it Swift}, respectively. Table \ref{TriggerTable} shows the used trigger constraints on ASM [2-12 keV] and BAT [15-50 keV] fluxes for each of the observed sources. Cyg X-3 was observed in 2007 for 28 hours; GRS 1915+105 for 14 hours in June 2007, and Cyg X-1 for 21 hours between July and November 2007. 
SS~433, was observed during the second half of August 2008 for 15
hours. The precession of the system can provoke a periodic (P$_{\Psi} \sim 162$~d) attenuation
 of the VHE gamma-rays produced near the compact object. We
observed the source following a predicted minimum of the opacity \cite{Reynoso2008},
avoiding also the eclipsing companion which cover the jet inner regions
each $\sim 13$ d.

\begin{table}[h]
\begin{center}

\caption{Trigger criteria for the microquasar observations in 2007}
\begin{tabular}{c|c|c|c}
\hline
\hline
& {\bf BAT daily} &{\bf BAT orbital} & {\bf ASM}\\
& [counts/s] & [counts/s] & [counts/s]\\
\hline
{\bf Cyg X-1} & / & $>$ 0.2 & $<$ 200 \\
{\bf Cyg X-3} & $>$ 0.05 & / & $<$ 200\\
{\bf GRS 1915+105}& $>$ 0.08 & / & $<$ 1000\\
\hline
\hline
\end{tabular}
\label{TriggerTable}
\end{center}
\end{table}
\vspace{-0.0cm}
\section{Search for steady and transient VHE emissions} \label{Magic}
\begin{figure}
\centering
\includegraphics[width=0.45\textwidth]{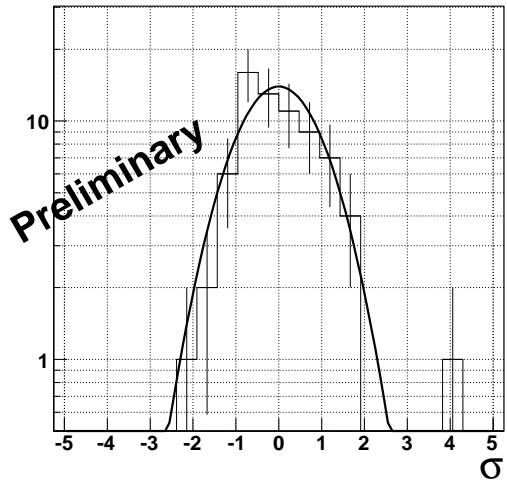}
\caption{Significance distribution of 70 night observations of Cyg X-1, Cyg X-3, GRS~1915+105 and SS~433. A Gaussian distribution with the mean of 0 and the standard deviation of 1 for 70 samples is also shown. The Cyg~X-1 TeV flare evidence appears at a 4.0 $\sigma$ level. \vspace{-0.5cm}}
\label{Significance}
\end{figure}

No signal of steady emission has been found from any of the four sources; in
Table \ref{steady} we report the integral upper limits on their VHE emission. Their
values
vary between 0.6 to 2.6\% C.U, and are calculated
for E $> 250$~GeV at a 95\% confidence level assuming a power law spectrum with a spectral index of $-2.6$. The differential upper limits are shown in Table~\ref{DiffULTable}.
To search for transient emission, data have been analysed also night by night. There are 28, 21, 16, and 5 nights of observations with good conditions for Cyg X-1, Cyg X-3, GRS~1915+105 and SS~433, respectively.
 The distribution of the significance of the 70 observations is shown in Fig.~\ref{Significance}. There is one observation with 4.0 $\sigma$ excess, which is Cyg X-1 observation mentioned above. Apart from it, the significances follow the normal Gaussian showing no signal evidence. The upper limit on VHE gamma-ray flux $>250$~GeV for individual night is $\sim 10$\% C.U.

Phase-folded analysis has also been performed (see Table~\ref{PBP}), to search for possible variations of the gamma-ray emission at different orbital phases.
No significant excess was found at any phase.

\section{Discussion} \label{Discussion}
\noindent VHE gamma-rays are expected to be produced in a MQ scenario. The emission should be naturally produced in the interaction between jet accelerated particles and the matter and radiation fields in the binary system and surrounding medium (see, e.g. \cite{Bosch-Ramon2009} $-$ \cite{Bordas2009}). The variability and the spectral properties in these sources can be however very complex, depending on multiple conditions and parameters largely unconstrained. To estimate the TeV emission from these systems one has to take into account their intrinsic capability to accelerate particles to multi-TeV energies, the emission outcomes through different radiation mechanisms and the absorption of the emerging VHE flux, which can be strong enough to attenuate most of the putative signal at a given orbital phase.


 \begin{table}[h]
   \caption{Steady emission flux integral upper limits above 250 GeV \vspace{-0.0cm}}
   \label{steady}
   \centering
   \begin{tabular}{|c|c|c|c|c|c|c|}
     \hline
       &  Eff.     &          &          & U.L. \\
       &  Time &          &          & [$10^{-12}$cm$^{-2}$s$^{-1}$] \\
       &  [h] & N$_{ex}$ & $\sigma$ &  / [C.U.]\\
     \hline 
     Cyg X-1& 36.2 & -49.6 $\pm$ 51.7 & -0.96 & 1.24 / 0.75\% \\  \hline
     Cyg X-3& 32.8 & -56.4 $\pm$ 47.1 & -1.19 & 1.01 / 0.61\% \\ \hline
     GRS~1915& 22.3 & -94.0 $\pm$ 52.3 & -1.87 & 1.17 / 0.71\% \\ \hline
     SS~433&  6.6 &  -9.7 $\pm$ 27.9 & -0.35 & 4.30 / 2.61\% \\
     \hline
   \end{tabular}
\vspace{-0.0cm}
 \end{table}

\begin{table*} 
 \caption{Differential upper limits for the steady VHE emission.} 
\label{DiffULTable}
\begin{tabular}{|c|c|c|c|c|c|c|c|}\hline 
\multicolumn{2}{|c|}{Energy [GeV]} & 100 -- 178 & 178 -- 316 & 316 -- 562 & 562 -- 1000 & 1000 -- 1780 & 1780 -- 3160 \\
\hline
 & Cyg X-1  & $1.73 \times 10^{-10}$  & $2.75 \times 10^{-11}$  & $3.95 \times 10^{-12}$  & $1.97 \times 10^{-12}$  & $4.22 \times 10^{-13}$  & $8.97 \times 10^{-14}$ \\ \cline{2-8} 
 U.L. & Cyg X-3  & $6.54 \times 10^{-10}$  & $4.64 \times 10^{-11}$  & $5.15 \times 10^{-12}$  & $1.06 \times 10^{-12}$  & $4.58 \times 10^{-13}$  & $1.07 \times 10^{-13}$ \\ \cline{2-8} 
 [cm$^{-2}$TeV$^{-1}$s$^{-1}$] & GRS 1915 & $2.49 \times 10^{-10}$  & $3.03 \times 10^{-11}$  & $4.57 \times 10^{-12}$  & $1.84 \times 10^{-12}$  & $4.57 \times 10^{-13}$  & $1.14 \times 10^{-13}$ \\ \cline{2-8} 
 & SS 433  & $6.30 \times 10^{-10}$  & $6.70 \times 10^{-11}$  & $7.60 \times 10^{-12}$  & $3.50 \times 10^{-12}$  & $1.90 \times 10^{-12}$  & $3.40 \times 10^{-13}$ \\ \hline 
 \end{tabular} 
 \end{table*} 

The observations presented in this work have been performed following a trigger strategy intended to ensure the {\it{low/hard}} state of the sources, where jets are thought to be present \cite{Fender2004} \cite{McClintock2006}. Furthermore, our results span a wide range of the phases in which each source can be found during its orbital motion. The phase-folded reported upper limits can then be used to constrain the VHE emission in conditions where absorption should be the minimum allowed in each case, though the precise attenuation factor will depend on the intensity of the companion and disk photon fields, the precise location of the emitter with respect to them and the interaction angle between the gamma-ray and the target photons (see, for instance, \cite{Bednarek2007} \cite{Bosch-Ramon2008}).

In a leptonic framework, Inverse Compton (IC) process might produce detectable VHE fluxes when scattering off external photon fields permeating the jet and surroundings. In addition, one should consider the contributions from IC on self-generated synchrotron photon fields as well as relativistic Bremsstrahlung, which could significantly enhance the total VHE output. 

In a hadronic scenario, gamma-ray fluxes could be generated through the decay of $\pi^{0}$ produced in the interactions of relativistic protons in the jet against thermal ions from the companion wind or within the jet itself. Fluxes up to few~\% of the jet kinetic luminosity could be radiated at VHE \cite{Reynoso2008} \cite{Orellana2007} \cite{Romero2003}.

The integral flux upper limits $\phi_{\rm E_{\gamma} \ge 250}$ for the central regions of the observed sources reported in Tables~\ref{steady} and \ref{PBP} could point to some physical mechanism(s) by which the putative VHE fluxes would be suppressed.
Synchrotron losses in the magnetic field present in the jets, for instance, can play a major role in the final spectral energy distribution of MQs. A high enough magnetic equipartition fraction $\eta_{\rm eq} \equiv u_{\rm mag}/u_{\rm kin}$, where $u_{\rm mag}$ and $u_{\rm kin}$ are the magnetic and kinetic energy densities in the jet, respectively, would make this channel dominate the radiative losses and the gamma-ray fluxes produced through IC processes would be reduced. Considering only the companion's radiation energy density $u_{\star}= L_{\star}/4\pi d^{2} c $ , where $d^2 =(a_{orb}^{2} + z_{0}^{2})$, being
$L_{\star}$ the companion luminosity,  $a_{orb}$ the orbital separation at
a certain phase and $z_{0}$ the emitter location along the jet axis, the non-detection of steady gamma-ray emission could imply $u_{\rm mag} > u_{\star}$, leading to values of $\eta_{\rm eq} > 1.9 \times 10^{-3} \times (L_{\star \,39}/L_{\rm kin \,37})(\psi / 0.5^{\circ})^{2}[z_{0}^2/(z_0^2+a_{\rm orb}^{2})]$, where $\psi$ is the jet semiaperture angle, $L_{\star \,39 }= L_{\star}/10^{39}$~erg~s$^{-1}$, and $L_{\rm kin \,37}=L_{\rm kin}/10^{37}$~erg s$^{-1}$ is the jet kinetic power.

\begin{table*}
\caption{Phase-folded integral upper limits above 250 GeV}
\label{PBP}
\begin{tabular}{|c|c|c|c|c|c|c|c|c|c|c|c|}\hline
 \multicolumn{2}{|c|}{Phase\footnotemark} & 0.0-0.1 & 0.1-0.2 & 0.2-0.3 & 0.3-0.4 & 0.4-0.5 & 0.5-0.6 & 0.6-0.7 & 0.7-0.8 & 0.8-0.9 & 0.9-1.0 \\
\hline 
\hline 
 & Effective Time [h] & 4.9 & 1.8 & 3.2 & 0.5 & 3.9 & 6.6 & 2.6 & 2.8 & 4.0 & 5.9 \\ \cline{2-12}
 Cyg X-1 & $\sigma$ & 0.4 & -0.6 & -0.3 & 0.5 & -1.0 & 0.0 & -2.9 & -0.6 & -1.3 & 1.4 \\ \cline{2-12}
Period: 5.6 d & U.L. [$10^{-12}$cm$^{-2}$s$^{-1}$] & 7.3 & 8.0 & 7.3 & 30 & 3.7 & 5.6 & 2.6 & 5.6 & 3.0 & 10 \\ \hline
\hline 
 & Effective Time [h] & 3.1 & 3.0 & 2.8 & 3.1 & 3.0 & 3.6 & 4.5 & 3.5 & 3.5 & 2.5 \\ \cline{2-12}
 Cyg X-3 & $\sigma$ & 0.1 & -1.0 & -0.2 & -1.2 & -1.4 & -0.7 & 1.3 & 1.1 & 0.6 & -0.4 \\ \cline{2-12}
 Period: 0.2 d& U.L. [$10^{-12}$cm$^{-2}$s$^{-1}$] & 7.1 & 3.7 & 6.0 & 3.6 & 3.4 & 5.0 & 13 & 13 & 11 & 5.8 \\ \hline \hline 
 & Effective Time [h] & 2.0 & 6.0 & 0.0 & 0.0 & 0.0 & 5.1 & 4.6 & 3.9 & 0.1 & 0.5 \\ \cline{2-12}
 GRS~1915 & $\sigma$ & -0.5 & -1.0 & -- & -- & -- & -1.9 & -0.2 & -0.9 & -0.9 & 1.0 \\ \cline{2-12}
Period: 33.5 d & U.L. [$10^{-12}$cm$^{-2}$s$^{-1}$] & 10 & 4.7 & --& --& --& 2.1 & 4.8 & 3.6 & 41 & 43 \\ \hline \hline 
 & Effective Time [h]           & 0.98 & 0.0 & 0.0 & 0.0 & 0.0 & 1.4 & 0.0  & 0.45 & 2.2 & 1.5 \\ \cline{2-12}
 SS~433 & $\sigma$     & -1.5  & --  & --  & --  & --  & 1.2 & --  & -0.24& 0.1 &-0.68 \\ \cline{2-12}
 Period: 13.1 d& U.L. [$10^{-12}$cm$^{-2}$s$^{-1}$]  & 6.5& --  & --  & --  & --  & 21&--& 18 & 9.8 & 7.7 \\ \hline

\hline
\end{tabular}
\vspace{-0.5cm}
\end{table*}

In addition, the gamma-ray luminosities depend linearly on $q_{\rm accel}$, the fraction of $L_{\rm kin}$ transferred to an accelerated e/p plasma, which is poorly constrained under a theoretical point of view. It
could be much less efficient than expected, though it seems unlikely since
acceleration is actually required to explain the persistent leptonic
non-thermal
emission at lower radio energies and probably at hard X-ray energies. Interestingly, Reynoso et
al.~\cite{Reynoso2008} have recently estimated the gamma-ray production in
the
hadronic jets of SS~433. By using the HEGRA upper limit on the VHE flux
from this
system \cite{HEGRA},
$q_{\rm accel}$ for protons is constrained to be $ \leq 3 \times 10^{-4}$.
These values assume however that HEGRA observations took place during unknown
precessional phases $\Psi$. In order to account for the absorption of the
gamma-ray photons, a $\Psi$-averaged flux was used to derive the upper
limit on
$q_{\rm accel}$. The current MAGIC observations were performed during the
lowest
absorption phases. By comparing the proton-proton expected gamma-ray
fluxes with
our upper limits, a more stringent constraint of the hadronic $q_{\rm accel}\le 7.4 \times
10^{-5}$ can be derived.

In addition to the steady emission, short-lived ejections can also occur in the transitions between the {\it{low/hard}} to the {\it{high/soft}} MQ states, although the flaring behaviour is not limited to the state changes in sources like GRS~1915~+105 or Cyg~X-3, and is quite unpredictable. The kinetic power expelled in such episodes can be up to several times 10$^{39}$ erg~s$^{-1}$ \cite{Atoyan1999}, producing intense radiation fluxes at all wavelengths. The appearance of outbursts at radio/IR wavelengths can be related to an increase of the activity also at VHE, although the time-delay between them could range from hours to days \cite{Atoyan1999}. The hard X-ray/TeV correlation should provide more tightening constraints, since the emission could be produced by the same particle population or with similar timescale. However, the multifrequency correlation could be present only at some stages of the flare \cite{Albert2007}, rendering the triggering of TeV outburst detections a quite difficult task. 
The non-detection of any transient signal could imply the high absorption present in the inner regions of these systems. 
The gamma-ray attenuation with the stellar and accretion disk photon fields could nonetheless distribute the luminosity to lower gamma-ray energies through electromagnetic cascades \cite{Orellana2007} \cite{Bednarek2007}. This would increase the fluxes at ranges where observatories like {\it{Fermi}} operate, providing the most simultaneous triggering of the putative VHE gamma-ray production in these systems. 

\footnotetext{For Cyg X-1, GRS 1915 and SS443, phase 0 is defined as the moment when the companion star is between the central compact object and the observer. For Cyg X-3 phase 0 is the moment when the compact object is between the companion star and the observer.}


To conclude, steady emission from GRS~1915~+105, Cyg X-1, Cyg X-3 and SS~433 remains undetected at VHE, although an evidence of TeV flare has been found in Cyg X-1. A complete, global description of the gamma-ray emission expected accounting both for leptonic and hadronic processes in the different sources and absorption conditions at each orbital phase in which they have been observed is beyond the scope of this work. Nevertheless, some relevant theoretical predictions can already be tested, while further observations at VHE energies with present ground-based Cherenkov telescopes and at lower gamma-ray energies with space missions like {\it {Fermi}}, will provide new information on the physics present in these galactic VHE laboratories.

\section*{Acknowledgments}

 We would like to thank the Instituto de Astrofisica de
 Canarias for the excellent working conditions at the
 Observatorio del Roque de los Muchachos in La Palma.
 The support of the German BMBF and MPG, the Italian INFN
 and Spanish MICINN is gratefully acknowledged.
 This work was also supported by ETH Research Grant
 TH 34/043, by the Polish MNiSzW Grant N N203 390834,
 and by the YIP of the Helmholtz Gemeinschaft.

\end{document}